8# A Novel Synchronous Reference Frame Frequency-Locked Loop

Xiangjun Quan, *Member, IEEE*, Qinran Hu *Member, IEEE*, Alex Q. Huang, *Fellow, IEEE,* Xiaobo Dou *Member, IEEE*, Zaijun Wu *Member, IEEE**Abstract—* **This letter proposes a new design of frequency-locked loop (FLL) which is based on synchronous (*dq*) reference frame instead of stationary (*αβ*) reference frame. First, a synchronous reference frame FLL (briefly called SRF-FLL$_0$) equivalent to the conventional FLL is proposed. Then the SRF-FLL$_0$ is improved by utilizing the phase error to acquire a better performance. The small-signal modeling and parameter tuning of the improved synchronous reference frame FLL (SRF-FLL) are presented. Finally, the theoretical analysis and experiment results verify the superiority and effectiveness of proposed SRF-FLL.**

*Index Terms—* Frequency-locked loop, complex filter, synchronization, inverter control.## I. INTRODUCTION

Grid synchronization techniques are important for inverter control in power applications. The most widely used grid synchronization techniques are phase-locked loops (PLLs) [1] and frequency-locked loops (FLLs) [2], [3]. Generally, PLLs are mainly implemented in synchronous (*dq*) reference frame, while FLLs are realized in stationary (*αβ*) reference frame.

In recent years, PLLs gain great development on their filtering performance such as grid disturbance rejection and dynamic response [1], however, the development of FLLs is relatively slow. The reason is mainly attributable to their different working frames by Saeed Golestan *et al.* [2]. Because designing the filter in *αβ* reference frame is more complicated than designing in *dq* reference frame due to the difficult small-signal modeling in *αβ* frame (it is hard to build small-signal model for the AC signals in *αβ* frame).

The most popular FLL techniques are mainly based on second-order generalized integrator (SOGI) [4] and reduced-order generalized integrator (ROGI) [5], [6]. From the perspective of state observing, FLLs require frequency error information which is easily acquired by the cross product of the grid voltage and its estimate in *αβ* frame. While for PLLs, the phase error information represented by the *q*-axis voltage is needed, hence PLLs are generally realized in *dq* frame although sometimes the filter is implemented in *αβ* frame [7]. The difficulty of modeling in *αβ* frame makes the tuning of FLLs more complicated, so that the filtering capability of FLLs is not well developed. In [4]-[6], a first-order model is established for the frequency estimation loop for the parameter tuning, while a second-order model is adopted in [8], [9]. Obviously, the first-order model cannot represent the dynamic of FLLs, hence the second-order model is deduced in [9] to describe the dynamic response of the frequency estimate with better accuracy. According to the second-order model in [9], the damping oscillation frequency (the real part of the characteristic root) is only depended on the parameter of the ROGI/SOGI but irrelevant with the frequency estimate gain. This introduces some difficulties for the parameter determination of FLLs.

To improve the filtering performance of FLLs, this letter proposes a new framework for FLLs which is implemented in synchronous (*dq*) reference frame. The proposed synchronous (dq) reference frame FLL (SRF-FLL) can utilize frequency error and phase error information simultaneously, so that an extra loop filter can be designed to enhance the performance of the FLL. Then a novel loop filter is designed to achieve a better performance without additional computation. Finally, the proposed SRF-FLL is tested by a TMS320F28379D based testbed, and the experiment results verify its improvements.

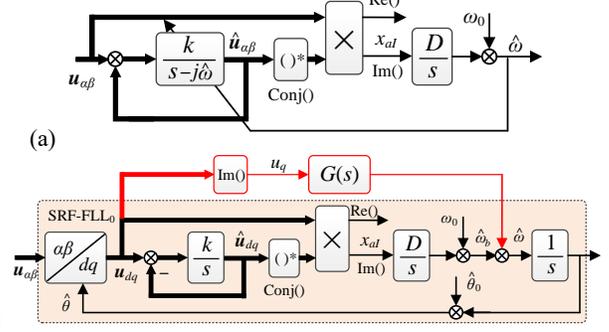

Fig. 1 (a) The conventional FLL and (b) the proposed SRF-FLL.

## II. NEW FLL IN SRF

### A. Equivalent SRF-FLL$_0$

According to [9], the conventional FLL is shown in Fig. 1 (a) where the bold faces and lines denote the complex variable. $\boldsymbol{u}_{\alpha\beta} = u_\alpha + ju_\beta$ and $\boldsymbol{u}_{dq} = u_d + ju_q$ are the grid voltage in *αβ* frame and *dq* frame respectively. The transfer function (TF) of the conventional FLL was reported in [9]

$$\hat{\omega} = \frac{DV^2}{s^2 + ks + DV^2}\omega \qquad (1)$$

Fig. 1 (b) shows the framework of the proposed SRF-FLL. First, the FLL in the dotted frame in Fig. 1 (b), referred as SRF-FLL$_0$, is investigated. The estimation of $\boldsymbol{u}_{dq}$ is expressed as

$$\dot{\hat{\boldsymbol{u}}}_{dq} = k(\boldsymbol{u}_{dq} - \hat{\boldsymbol{u}}_{dq}) \qquad (2)$$

which is a complex low pass filter (LPF). This LPF can be derived by transforming the ROGI into *dq* frame. It is used here to acquire the frequency error information. Considering the fundamental positive sequence voltage

$$\boldsymbol{u}_{\alpha\beta} = u_\alpha + jv_\beta = V\cos\theta + jV\sin\theta = Ve^{j\theta} \qquad (3)$$

where $\theta = \omega t + \theta_0$, the voltage in *dq* frame is

$$\boldsymbol{u}_{dq} = \boldsymbol{u}_{\alpha\beta}e^{-j\hat{\theta}} = Ve^{j\theta - j\hat{\theta}} = Ve^{j\theta_e} = V\cos\theta_e + jV\sin\theta_e \qquad (4)$$

where $\hat{\theta} = \hat{\omega} t + \hat{\theta}_0$ and $\theta_e = \theta - \hat{\theta}$, $\hat{\omega}$ is the frequency estimate, $\hat{\theta}$ is the generated phase which may not exactly equal $\theta$, $\hat{\theta}_0$ is the initial phase of $\hat{\theta}$. An auxiliary variable is defined as $\boldsymbol{x}_a = \boldsymbol{u}_{dq}\hat{\boldsymbol{u}}_{dq}^* = x_{aR} + jx_{aI}$ where "*" represents complex conjugate. The dynamic of $\boldsymbol{x}_a$ is

$$\begin{aligned}\dot{\boldsymbol{x}}_a &= \dot{\boldsymbol{u}}_{dq}\hat{\boldsymbol{u}}_{dq}^* + \boldsymbol{u}_{dq}\dot{\hat{\boldsymbol{u}}}_{dq}^* \\ &= j\omega_e\boldsymbol{u}_{dq}\hat{\boldsymbol{u}}_{dq}^* + \boldsymbol{u}_{dq}\left(k\boldsymbol{u}_{dq} - k\hat{\boldsymbol{u}}_{dq}\right)^* \\ &= j\omega_e\boldsymbol{x}_a - k\boldsymbol{x}_a + kV^2\end{aligned} \qquad (5)$$

where $\omega_e = \omega - \hat{\omega}$ is the frequency error. In (5) the relationship $\dot{\boldsymbol{u}}_{dq} = j\dot{\theta}_e V e^{j\theta_e} = j\omega_e\boldsymbol{u}_{dq}$ is used. Equation (5) is also a complex LPF whose settling time is only depended on parameter $k$, while $\omega_e$ just affects the oscillation part during the dynamic process and $\omega_e \ll k$. The steady-state solution of (5) is

$$\bar{x}_a = \bar{x}_{aR} + j\bar{x}_{aI} = \frac{k^2 + jk\omega_e}{k^2 + \omega_e^2}V^2 \quad (6)$$

The imaginary part of the auxiliary variable includes the frequency error, hence, like the conventional FLL, $x_{aI}$ in $dq$ frame can also be used to estimate the frequency as shown in Fig. 1:

$$\hat{\omega} = \frac{D}{s}x_{aI} \quad (7)$$

Differently, in conventional FLL, the auxiliary variable is defined as $x_a = u_{\alpha\beta}\hat{u}_{\alpha\beta}^*$ as shown in Fig. 1 (a). Considering the condition of $\omega_e \ll k$, the dynamic of the frequency error transmission can be described by

$$\dot{x}_{aI} = \omega_e \bar{x}_{aR} - kx_{aI} \approx \omega_e V^2 - kx_{aI} \quad (8)$$

which means $\frac{\Delta x_{aI}}{\Delta\omega_e} = \frac{V^2}{s+k}$. Then the small-signal block diagram is derived as shown by the dotted frame in Fig. 2 where the prefix "$\Delta$" denotes the small signal. If the frequency estimate gain is selected as

$$D = \frac{k}{V^2}d \quad (9)$$

which is a conventional selection in most literatures. The closed-loop TF can be acquired

$$\Delta\hat{\omega} = \frac{DV^2}{s^2 + ks + DV^2}\Delta\omega = \frac{kd}{s^2 + ks + kd}\Delta\omega \quad (10)$$

This model is identical with (1). It is demonstrated that the proposed SRF-FLL$_0$ is the $dq$-frame version of the conventional FLL. In this condition, $u_q$ may not be zero. The phase difference between the grid voltage and the generated phase is computed as

$$\hat{\theta}_e = \arctan(\hat{u}_q/\hat{u}_d) \quad (11)$$

It can be adjusted by change $\hat{\theta}_0$. Then the grid phase is acquired as

$$\hat{\theta}_{est} = \hat{\theta} + \hat{\theta}_e \quad (12)$$

and the estimate of $u_{dq}$ orientated to grid voltage is obtained as

$$\hat{u}_{dqest} = u_{\alpha\beta}e^{-j\hat{\theta}_{est}} = u_{\alpha\beta}e^{-j(\hat{\theta}+\hat{\theta}_e)} = u_{dq}e^{-j\hat{\theta}_e} = Ve^{j\theta_e - j\hat{\theta}_e} = V + j0 \quad (13)$$

### B. Improved SRF-FLL

Different from the conventional FLL, the SRF-FLL$_0$ has an advantage that the phase error can be acquired by $u_q$. Therefore, the improved SRF-FLL is upgraded as

$$\hat{\omega} = \frac{D}{s}x_{aI} + G(s)u_q \quad (14)$$

where $G(s)$ is the extra loop filter. Supposing $\theta_e \approx 0$ (this condition can be acquired by tuning $\hat{\theta}_0$), the small-signal model of the frequency estimation (14) can be acquired as shown in Fig. 2. The open-loop gain of the frequency estimation loop is

$$G_{of} = \frac{d}{s}\left(\frac{k}{s+k} + \frac{V}{d}G(s)\right) \quad (15)$$

From (15) the open-loop gain can be adjusted by designing the extra loop filter $G(s)$. This offers abundant additional degree of the freedom (DOF) for the FLL technique.

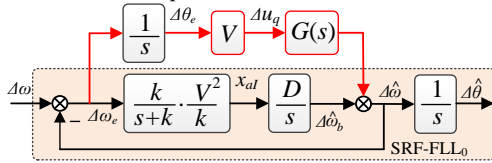

Fig. 2 The small-signal model of the proposed SRF-FLL.

### III. ONE DESIGN OF $G(s)$

The selection of $G(s)$ is flexible, large number of filters could be adopted for some specific targets. For example, the PI controller could be adopted, then $u_q$ will be controlled to zero. In this condition, the SRF-FLL is a combination of PLL and FLL (SRF-FLL-PLL). The assumption of $\theta_e \approx 0$ is automatically satisfied. However, the performance needs further study. This letter will not focus on the SRF-FLL-PLL but give a new design of $G(s)$.

### A. SRF-FLL with Selected $G(s)$

Observing (15), it can be found if $G(s)$ is selected as

$$G(s) = \frac{d}{V}\frac{s}{s+k} \quad (16)$$

the open-loop gain of the FLL becomes

$$G_{of} = \frac{d}{s}\left(\frac{k}{s+k} + \frac{s}{s+k}\right) = \frac{d}{s} \quad (17)$$

and correspondingly, the closed-loop TF of the FLL is evolved as

$$\Delta\hat{\omega} = \frac{d}{s+d}\Delta\omega \quad (18)$$

which is a first-order LPF. If choosing $\hat{\omega}_b$ as the frequency estimate, the model can be derived from Fig. 2

$$\Delta\hat{\omega}_b = \frac{k}{s+k}\frac{d}{s+d}\Delta\omega = \frac{kd}{s^2 + (k+d)s + kd}\Delta\omega \quad (19)$$

Comparing to (1) or (10), the damping oscillation frequency is improved by the frequency estimation gain $d$.

Moreover, according to (2), the output of $G(s)$ can be derived directly from the error of $u_q$:

$$\frac{d}{V}e_q = \frac{d}{V}(u_q - \hat{u}_q) = \frac{d}{V}\frac{s}{s+k}u_q = G(s)u_q \quad (20)$$

Consequently, the realization of $G(s)$ is avoided as shown in Fig. 3.

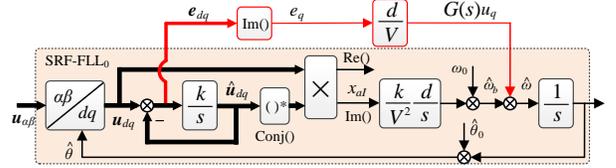

Fig. 3 The proposed SRF-FLL with selected $G(s)$.

### B. Phase Analysis

The phase transfer feature of the proposed SRF-FLL is different from the conventional FLL. First the model of the generated phase can be derived from (18)

$$\Delta\hat{\theta} = \frac{d}{s+d}\Delta\theta \quad (21)$$

Second, for the estimate of the phase difference $\hat{\theta}_e$, differentiating (11) with respect to time gives

$$\dot{\hat{\theta}}_e = \frac{\dot{\hat{u}}_q\hat{u}_d - \hat{u}_q\dot{\hat{u}}_d}{\hat{u}_d^2 + \hat{u}_q^2} = \frac{k}{V^2}(\hat{u}_d u_q - \hat{u}_q u_d) = \frac{k}{V^2}x_{aI} \quad (22)$$

Combining (8) and (22) gives

$$\Delta\hat{\theta}_e = \frac{k}{s+k}\frac{1}{s}\Delta\omega_e = \frac{k}{s+k}\Delta\theta_e = \frac{ks}{(s+k)(s+d)}\Delta\theta \quad (23)$$

Finally, Substituting (21) and (23) into (12) yields

$$\Delta\hat{\theta}_{est} = \frac{ks + ds + kd}{(s+k)(s+d)}\Delta\theta = \frac{(k+d)s + kd}{s^2 + (k+d)s + kd}\Delta\theta \quad (24)$$

which describes the phase transfer feature of the proposed SRF-FLL.

### IV. EXPERIMENTAL ANALYSIS AND VERIFICATION

#### A. Performance Comparison

To clearly demonstrate the advantage of the proposed SRF-FLL, the model, analysis and design guideline are summarized in TABLE I. The SRF-FLL$_0$ is omitted, since it is equivalent to the conventional FLL. The selection of $k$ affects the filter performance of ROGI and LPF which determines the suppression ability of the harmonics if there are no multiple ROGIs are configured, hence, generally, $k$ is tuned according to the filter requirement of ROGI or LPF. Then, for the conventional FLL, $d$ only changes the damping factor $\zeta$, but cannot adjust damping oscillation frequency $\zeta\omega_n$. This implies that the regulation ability of $d$ is limited. It has an optimal design of $d=0.5k$ resulting in $\zeta = 0.707$ [9]. While the proposed SRF-FLL always has two real roots which means that $d$ can improve the damping oscillation frequency without deterioration of damping. The Bode plots for the conventional FLL with $d = k$ and $d = 0.5k$ (the optimal design) and the proposed SRF-FLL ( $\hat{\omega}_b$ is the estimate) with $d = k$ are displayed. As can be observed in Fig. 4, the proposed SRF-FLL has better filter ability with the same $k$, though the damping oscillation frequency of the SRF-FLL is double of its counterpart in the conventional FLL.

### B. Performance Verification

To verify the proposed SRF-FLL, a testbed using a TMS320F28379D-based 32-bit floating-point DSP at 200 MHz was constructed. The peak value of the voltage is nominalized to 1 with frequency 60 Hz. Since the proposed method mainly focuses on the FLL design for the fundamental voltage, the fundamental positive sequence component is used for the test. If the input voltage is polluted by harmonics, the multiple prefilters (multiple SOGIs, ROGIs, etc. [5][6]) can be used to address the effects of the harmonics and imbalance, thus the performance of the proposed method will not be affected.

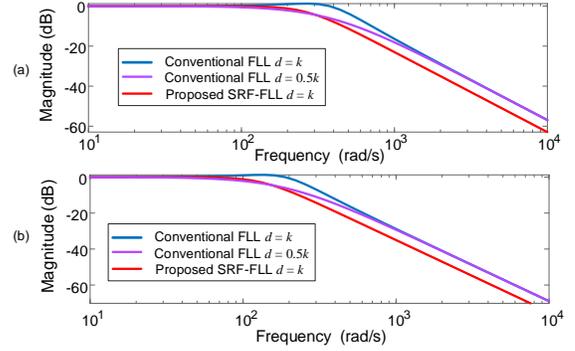

Fig. 4 Closed-loop magnitude frequency response of the estimation of frequency: (a) $k = 120\pi$, (b) $k = 60\pi$.

TABLE I COMPARISON OF CONVENTIONAL FLL AND PROPOSED SRF-FLL

| | | $\Delta\hat{\omega}$ - $\Delta\omega$ TF | $\Delta\hat{\theta}$ - $\Delta\theta$ TF | Roots | Root analysis and Design guideline |
|---|---|---|---|---|---|
| Conventional FLL | | $\dfrac{kd}{s^2 + ks + kd}$ | $\dfrac{ks + kd}{s^2 + ks + kd}$ | $p = \dfrac{-k \pm \sqrt{k^2 - 4kd}}{2}$ $= -\zeta\omega_n \pm \omega_n\sqrt{\zeta^2 - 1}$ | $\zeta\omega_n = 0.5k$, the time constant is $0.5k$ $\zeta : \begin{cases} \zeta \geq 1 & d \leq 0.25k \\ 1/\sqrt{2} \leq \zeta < 1 & 0.25k < d \leq 0.5k \\ \zeta < 1/\sqrt{2} & d > 0.5k \end{cases}$ 1), choosing $k$ with comprehensive consideration of the filter ability of the ROGI shown in Fig. 1 (a) and the settling time. 2), the optimal selection $d=0.5k$ resulting in $\zeta = 0.707$ [9]. |
| Proposed SRF-FLL | $\Delta\hat{\omega}_b$ | $\dfrac{kd}{s^2 + (k+d)s + kd}$ | $\dfrac{(k+d)s + kd}{s^2 + (k+d)s + kd}$ | $p_1 = -k, p_2 = -d$ | $\zeta \geq 1$, $\zeta\omega_n = 0.5(d+k)$, the time constant is $d$ and $k$. 1), choosing $k$ just considering the filter ability of the LPF (2). 2), the settling time depends on $d$ and $k$. 3), the best choice is to design $d=k$. |
|  | $\Delta\hat{\omega}$ | $\dfrac{d}{s + d}$ | | $p_1 = -d$ | |

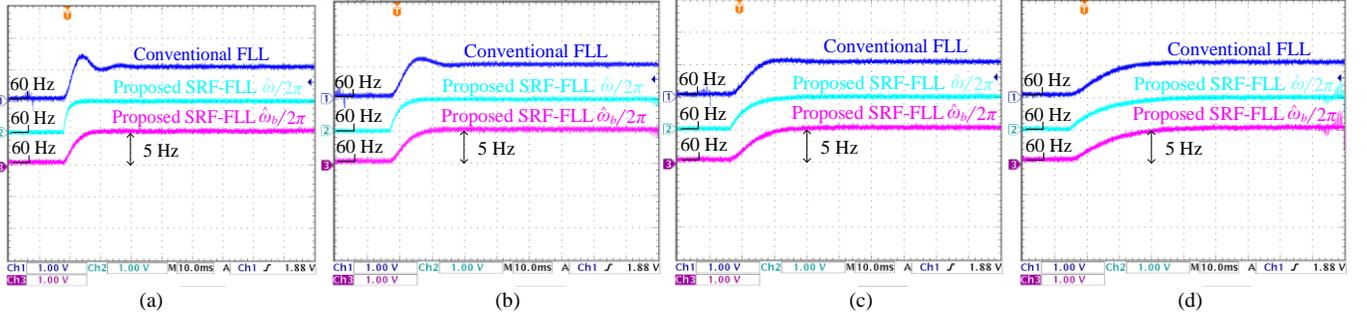

Fig. 5 Frequency step (+5 Hz) response comparison with $k = 120\pi$: (a) $d = 2k$; (b) $d = k$; (c) $d = 0.5k$; (d) $d = 0.25k$.

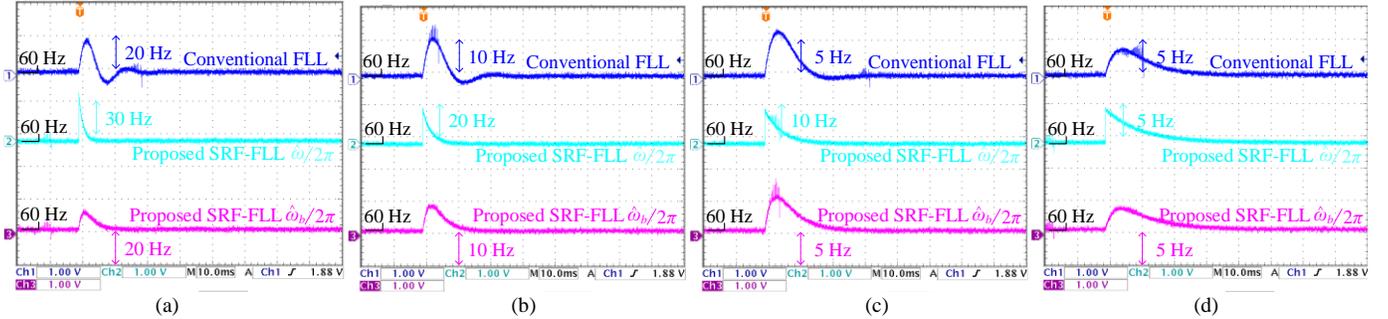

Fig. 6 Phase step (20 degree) response comparison with $k = 120\pi$: (a) $d = 2k$; (b) $d = k$; (c) $d = 0.5k$; (d) $d = 0.25k$.

First, according to the design guideline shown in TABLE I, the proposed SRF-FLL and the conventional FLL are compared under the conditions: $k = 120\pi$, and (a) $d = 2k$; (b) $d = k$; (c) $d = 0.5k$; (d) $d = 0.25k$. The selection of $k$ has no inference for the comparison results, since according to the magnitude frequency response in Fig. 4, the effect of $k$ is same for both FLLs. In addition, the same $k$ results to the same filter ability for the LPF in the proposed SRF-FLL as well as the ROGI in conventional FLL. Therefore, it is fair to make the comparison with same $k$. $d$ is chosen according to the critical point of $\zeta$. Design of $d = 2k$ is added to verify the effect of larger $d$.

From Fig. 5, it can be observed that with the same $k$, the proposed SRF-FLL performs better when the grid frequency steps up 5 Hz. With the increasing of $d$, the dynamic response of the conventional FLL deteriorates, while the proposed SRF-FLL always manifests good dynamic response either for $\hat{\omega}$ or $\hat{\omega}_b$. This lies in the damping factor of the conventional FLL decreasing with the increase of $d$ as shown in TABLE I. However, for the proposed SRF-FLL, the behavior of $\hat{\omega}$ is a first-order system as shown by (18) and, for $\hat{\omega}_b$, its damping is always more than 1. The same results can also be observed in Fig. 6 when the grid phase jumps 20 degree. Moreover, the results of $d = 2k$ provide the verification that larger $d$ can improve the regulation speed of the FLL without deterioration of damping factor. These results are coincident with the theoretical analysis. In addition, it can be seen that although $\hat{\omega}$ has the best performance with frequency step, it suffers large frequency variation when phase steps, this is because the error of $u_q$ is imposed directly to $\hat{\omega}$. Comparatively, $\hat{\omega}_b$ performs best when the phase changes. Hence, $\hat{\omega}_b$ can be selected as the final estimate of the frequency.

Moreover, $d=0.5k$ ensures an optimum design for the conventional FLL. Hence, it is fair to compare this optimal design with the proposed SRF-FLL, these two results are shown in Fig. 7. The fastest settling time is achieved by $\hat{\omega}$, while $\hat{\omega}_b$ and the conventional FLL have similar settling time, however, the conventional FLL performs a small overshoot due to the damping ratio of 0.707.

Second, to further verify the effectiveness of the proposed SFR-FLL, the conditions of amplitude, frequency and phase jumps are tested, and the estimations of $u_d$, $u_q$ (phase error) and frequency are monitored. The parameters are tuned as $d = k = 120\pi$. The results, shown in Fig. 8, prove that the proposed SRF-FLL performs well. The estimates of $u_d$, $u_q$ and $\omega$ manifest a good dynamic response when the amplitude, frequency and phase change. Fig. 8 (b) shows when frequency changes, the phase difference has a little variation. This accords with (23). Furthermore, Fig. 8 (c) verifies the phase estimate ability. When phase steps, the curve of $u_q$ demonstrates that the phase step is tracked successfully. Furthermore, the estimate $\hat{u}_{dqest}$ is also monitored when frequency and phase step as shown in Fig. 9. The results manifest that $\hat{u}_{dqest}$ successfully synchronizes to the grid voltage with zero-phase difference.

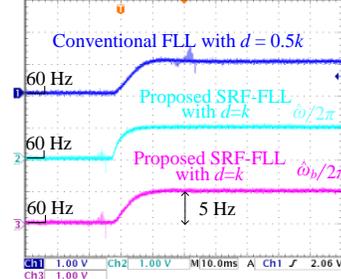

Fig. 7 Frequency step comparison to conventional FLL with the optimal design.

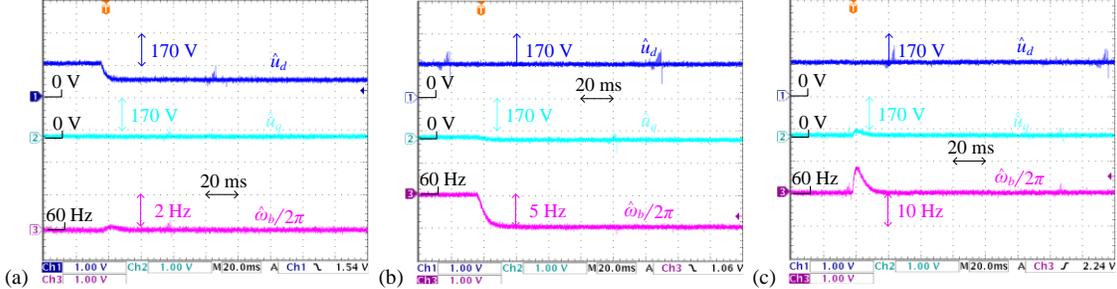

Fig. 8 The performance of the SRF-FLL with $d = k = 120\pi$: (a) amplitude sag (b) –5Hz frequency step; (c) 20-degree phase step.

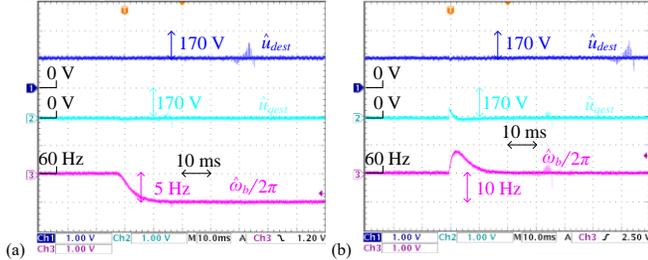

Fig. 9 Measurement of $\hat{u}_{dqest}$: (a) frequency step, (b) 20-degree phase step

## V. CONCLUSION

This letter presents a novel synchronous ($dq$) reference frame FLL (SRF-FLL). Comparing with the conventional FLL, the proposed SRF-FLL can acquire and utilize not only frequency error but also phase error. The theoretical analysis and experiment results demonstrate that the proposed SRF-FLL achieves better filter performance than conventional FLLs with the same prefilter parameter $k$. In proposed SRF-FLL, increasing the frequency estimate gain will not deteriorate the damping of the system, hence the better dynamic response is achieved. Further, the framework of the proposed SRF-FLL provides new insights for the design of FLLs by introducing extra loop filter $G(s)$. In the future, the authors will keep optimizing $G(s)$ for better performance.

REFERENCES

[1] S. Golestan, J. M. Guerrero, and J. C. Vasquez, "Three-phase PLLs: A review of recent advances," *IEEE Trans. Power Electron.*, vol. 32, no. 3, pp. 1894–1907, Mar. 2017

[2] S. Golestan, J. M. Guerrero, J. C. Vasquez, A. M. Abusorrah and Y. Al-Turki, "A Study on Three-Phase FLLs," in *IEEE Transactions on Power Electronics*, vol. 34, no. 1, pp. 213-224, Jan. 2019.

[3] Z. Xin, R. Zhao, P. Mattavelli, P. C. Loh, and F. Blaabjerg, "Re-investigation of generalized integrator based filters from a first-order-system perspective," *IEEE Access*, vol. 4, pp. 7131–7144, 2016.

[4] P. Rodríguez, A. Luna, I. Candela, R. Mujal, R. Teodorescu, and F. Blaabjerg, "Multiresonant frequency-locked loop for grid synchronization of power converters under distorted grid conditions," *IEEE Trans. Ind. Electron.*, vol. 58, no. 1, pp. 127–138, Jan. 2011.

[5] S. Vazquez, J. A. Sanchez, M. R. Reyes, J. I. Leon, J. M. Carrasco, "Adaptive vectorial filter for grid synchronization of power converters under unbalanced and/or distorted grid conditions," *IEEE Trans. Ind. Electron.*, vol. 61, no. 3, pp. 1355–1367, Mar. 2014.

[6] X. Quan, X. Dou, Z. Wu, M. Hu, and A. Q. Huang, "Complex-coefficient complex-variable-filter for grid synchronization based on linear quadratic regulation," *IEEE Trans. Ind. Informat.*, vol. PP, no. 99, pp. 1–1, 2017.

[7] X. Guo, W. Wu, and Z. Chen, "Multiple-complex coefficient-filter-based phase-locked loop and synchronization technique for three-phase grid-interfaced converters in distributed utility networks," *IEEE Trans. Ind. Electron.*, vol. 58, no. 4, pp. 1194–1204, Apr. 2011.

[8] S. G. Jorge, C. A. Busada, and J. A. Solsona, "Frequency adaptive discrete filter for grid synchronization under distorted voltages," *IEEE Trans. Power Electron.*, vol. 27, no. 8, pp. 3584–3594, Aug. 2012.

[9] S. Golestan, J. M. Guerrero and J. C. Vasquez, "High-Order Frequency-Locked Loops: A Critical Analysis," in *IEEE Transactions on Power Electronics*, vol. 32, no. 5, pp. 3285-3291, May 2017.